\documentclass[showpacs,twocolumn,aps]{revtex4}
\usepackage{amsmath,amssymb}
\begin{document}
\def\cs#1#2{#1_{\!{}_#2}}
\def\css#1#2#3{#1^{#2}_{\!{}_#3}}
\def\ket#1{|#1\rangle}
\def\bra#1{\langle#1|}

\title{Testing Quantum Gravity via Cosmogenic Neutrino Oscillations}

\author{Joy Christian}

\email{joy.christian@wolfson.oxford.ac.uk}

\affiliation{Wolfson College, Oxford University, Oxford OX2 6UD, United Kingdom}

\date{20 September 2004}

\begin{abstract}
Implications of some proposed theories of quantum gravity for neutrino flavor oscillations
are explored within the context of modified dispersion relations of special relativity. In
particular, approximate expressions for Planck-scale-induced deviations from the standard
oscillation length are obtained as functions of neutrino mass, energy, and propagation distance.
Grounding on these expressions, it is pointed out that, in general, even those deviations that
are suppressed by the second power of the Planck energy may be observable for ultra-high-energy
neutrinos, provided they originate at cosmological distances. In fact, for neutrinos in
the highest energy range of EeV to ZeV, deviations that are suppressed by as much as the
seventh power of the Planck energy may become observable. Accordingly, realistic possibilities
of experimentally verifying these deviations by means of the next generation neutrino
detectors---such as IceCube and ANITA---are investigated.
\end{abstract}


\maketitle

\section{Introduction}

Despite many decades of intense efforts, the task of constructing a viable theory of quantum
gravity remains largely a speculative enterprise. There is, of course, no shortage of approaches
to quantum gravity, many with unprecedented mathematical sophistication and conceptual
innovation, but more often than not they harbor mutually incompatible philosophies
\cite{Rovelli-2000}. Worse still, the minuteness of the Planck length---or, equivalently, the
enormity of the Planck energy---guarantees to frustrate any attempt to experimentally
distinguish the better approaches from the worse. Amidst this dire state of affairs, in recent
years a glimmer of hope has emerged, and blossomed into a sub-discipline of ``the phenomenology
of quantum gravity'' \cite{Amelino-Camelia-2005}. This is based on the observation that,
regardless of incompatibility in philosophies and diversity in theoretical details, several
approaches to quantum gravity predict energy-momentum relations for elementary particles---in
the semiclassical or effective theory limit---that differ from their special relativistic
counterpart in somewhat similar manner
\cite{Samuel-1989,Ellis-2000a,Pullin-1999,Alfaro-2000,Alfaro-2002,Lehnert-2003,Bertolami-2004}. 
In natural units, these modified dispersion relations can be expressed as:
\begin{equation}
p^2+m^2 = E^2\left[1-\sum_{n=1}^{\infty}\,\xi^{(n)}\frac{\,E^n}{\,\css mnP}
\right]\!,\label{first}
\end{equation}
where ${\cs mP}$ is the Planck mass, and ${\xi^{(n)}}$ are dimensionless parameters, which
do not necessarily vanish for all orders of suppression by the Planck energy, and depend in
general on spin and helicity of the particles
\cite{Smolin-2003}. Clearly, away from the Planck regime
(i.e., for ${E\ll{\css m{}P}}$) these generalized relations effectively reproduce the familiar
special relativistic dispersion relation: ${E^2=p^2+m^2}$.

Remarkably, despite the suppressions by the Planck energy, it was pointed out in
Ref.\cite{Amelino-Camelia-1998} that delays in the times of arrival induced by these modified
relations in photons originating at cosmological distances may be observable, at
least up to the first-order of suppression. As anthologized in Ref.\cite{Amelino-Camelia-2005},
since then much effort has been devoted to understanding the phenomenological implications of
the relations such as (\ref{first}). For example, based on observations of synchrotron radiation
from the Crab nebula, together with a view that the modifications in (\ref{first}) result from
the existence of a preferred frame, strong bounds (of order ${10^{-9}}$) on the parameter
${\xi^{(1)}}$ have been obtained in Ref.\cite{Jacobson-2003} (see also
\cite{Jacobson-2003b,Stecker-2004,Jackiw-1990,Mewes-2001}).
Thus, it appears that theories of quantum gravity
that lead to a preferred frame, and as a result predict relations (\ref{first}) with the
parameter ${\xi^{(1)}}$ of order unity, have been ruled out by these observations. Moreover, it
has been suggested in Ref.\cite{Amelino-Camelia-2003} that advance neutrino observatories
such as ANTARES \cite{ANTARES} may well provide enough sensitivity to put significant bounds
even on the parameter ${\xi^{(2)}}$ in the relations (\ref{first}), thereby constraining those
theories of quantum gravity that predict only second and higher-order suppressions by the Planck
energy (see also section 5.4 of Ref.\cite{Jacobson-2004}).

Suppose now that, instead of simply providing constraining bounds, eventually a genuine
departure from the special relativistic dispersion relation is actually detected. Of course,
that would be a tremendous boost for some of the approaches to quantum gravity, not to mention
the much anticipated revolution it would finally bring about in physics. Inevitably, however,
such a detection would also leave a great deal of ambiguity as to which of the proposed
approaches to quantum gravity is truly compatible with the observed departure. For, as
mentioned above, several approaches lead to essentially the same quantitative predictions for
a departure from the special relativistic dispersion relation. It is therefore necessary to
search for options other than those existing in the literature, particularly those that are more
quantum mechanical in character. It is with this in mind that we explore here the implications
of the relations (\ref{first}) for neutrino flavor oscillations. To be sure, such a study has
been initiated before within specific models
\cite{Lambiase-2003,Brustein-2002,Ellis-2000,Bertolam-2000,Alfaro-2000,Barger-2000,Bahcall-2002,Choubey-2003,Mewes-2004}.
Our aim here, however, is to investigate the generic relations (\ref{first}), independently of
any specific model. Moreover, since the linear suppression by the Planck energy seems to have
been almost certainly ruled out \cite{Jacobson-2003}, our concern here would be to point out
that, in general, even those modified features of the standard theory of neutrino oscillations
that are quadratically suppressed by the Planck energy may be observable for ultra-high-energy
neutrinos, provided they have originated at cosmological distances. In fact, we shall show that,
for such ``cosmogenic'' neutrinos in the highest energy range of ${10^{18}}$ to ${10^{21}}$
eV, modifications that are as minute as
{\it cubically, quartically, quintically, sextically}, or even
{\it septically} suppressed by the Planck energy may also become observable. Accordingly, we
shall discuss realistic possibilities of verifying these modifications by means of the next
generation neutrino detectors, such as IceCube \cite{IceCube} and ANITA \cite{ANITA}.

To this end, in Section II we begin by briefly reviewing the standard theory of neutrino flavor
oscillations, thereby stressing our preferred prescription (as opposed to the one widespread
in textbooks) for arriving at the standard oscillatory phase factor. Then, in Section III, we
evaluate the Planck scale modifications of the standard theory, induced by the generalized
relations (\ref{first}), and by a related relation proposed in Ref.\cite{Christian-2004}.
Finally, before concluding in Section VI, we take up the issue of observability of these
modifications in Sections IV and V.

\section{The standard theory of neutrino flavor oscillations}

The remarkable and quintessentially quantum phenomena of neutrino flavor oscillations are
the result of the fact that neutrinos of definite flavor states ${\ket{\nu_{\alpha}}}$,
${\alpha=e,\mu,}$ or ${\tau}$, are {\it not} particles of definite mass states
${\ket{{\nu_j}}}$, ${j=1,2,}$ or ${3}$. Rather, they are coherent superpositions of the
definite mass states:
\begin{equation}
\ket{\nu_{\alpha}}=\sum_j U^*_{\alpha j}\,\ket{\nu_j},\label{super}
\end{equation}
with ${U}$ being the (time-independent) leptonic mixing matrix. By the same token, neutrinos
of definite mass states are coherent superpositions of the definite flavor states:
\begin{equation}
\ket{\nu_j}=\sum_{\beta} U_{\beta j}\,\ket{\nu_{\beta}},\label{flavour}
\end{equation}
with the mixing matrix being subject to the unitarity constraint
\begin{equation}
\sum_j U^*_{\alpha j} U_{\beta j}=\delta_{\alpha\beta}.\label{uni}
\end{equation}
(The reference we shall largely follow here is Ref.\cite{Kayser-2001}, but see also
Ref.\cite{Beuthe-2003} for a comprehensive review of the standard theory.) As a neutrino of
definite flavor state ${\ket{\nu_{\alpha}}}$---originating, say, from a distant cosmological
source---propagates through vacuum for a sufficiently long laboratory time, the heavier
mass-eigenstates in the superposition, such as (\ref{super}), lag behind the lighter ones, and,
as a result, the neutrino ends up arriving at a terrestrial detector in an altogether different
flavor state, say ${\ket{\nu_{\beta}}}$. The probability for this transition from one flavor
state to another can be easily obtained as follows. In the rest frame of each
${\ket{\nu_j}}$, where the proper time is ${\tau_j}$, plane wave analysis leads to the
Schr\"odinger equation
\begin{equation}
i\frac{\partial\;}{\partial\tau_j}\ket{\nu_j(\tau_j)}=m_j\,\ket{\nu_j(\tau_j)},\label{dynamics}
\end{equation} 
with a solution
\begin{equation}
\ket{\nu_j(\tau_j)}=e^{-im_j\tau_j}\ket{\nu_j(0)},\label{solution}
\end{equation}
where ${m_j}$ is the eigenvalue of the mass-eigenstate ${\ket{\nu_j(0)}}$. For our purposes it
is important to note that the phase factor ${e^{-im_j\tau_j}}$ in the last equation is
manifestly Lorentz invariant. In terms of the coordinate time ${t}$ and position ${\vec x}$ in
the laboratory frame, this phase factor takes the familiar form
\begin{equation}
e^{-i(E_jt-{\vec p}_j\cdot{\vec x})},\label{phase}
\end{equation}
where ${E_j}$ and ${{\vec p}_j}$ are, respectively, the energy and momentum associated with
the definite mass state ${\ket{\nu_j(0)}}$.

Now, neutrinos are highly relativistic
particles---i.e., they propagate with speeds extremely close to the speed of light, which
permits the convenient assumption ${t\approx x}$
${=L}$, where ${L}$ is the distance traversed by
neutrinos between production and detection. Moreover, assuming that the neutrinos are produced
with the same energy ${E}$ regardless of which state ${\ket{\nu_j(0)}}$ they are in (and that
${m_j\ll E}$), up to the second order in ${m_j}$ the special relativistic dispersion relation
gives the following expression for their momenta,
\begin{equation}
p_j=\sqrt{E^2-m_j^2}\,\approx E-\frac{\,m^2_j\,}{2E\,},\label{momen}
\end{equation}
which, along with the assumption ${t\approx L}$, reduces the phase factor in (\ref{phase}) to
\begin{equation}
e^{-i\frac{\,m^2_j}{\,2E\,}L}.
\label{approx-phase}
\end{equation}
Here the assumed equality of energy for all ${\ket{\nu_j(0)}}$ may seem to go against Lorentz
invariance, but it can be justified rigorously by noting that the quantum coherence we seek
would be maintained only between wavepacket components associated with the same energy
\cite{Stodolsky-1998}. Consequently, in the laboratory frame, and up to the second order in
${m_j}$, the time evolution of the neutrino flavor state (\ref{super})---duly respecting the
Lorentz invariance \cite{Giunti-2004}---is given by
\begin{equation}
\begin{split}
\ket{\nu_{\alpha}(t)}&=\sum_j U^*_{\alpha j}\,e^{-i\frac{\,m^2_j}{\,2E\,}L}\,\ket{\nu_j(0)} \\
&=\sum_{\beta}\sum_j U^*_{\alpha j}\,e^{-i\frac{\,m^2_j}{\,2E\,}L}\,U_{\beta j}\,
\ket{\nu_{\beta}(0)}.\label{super-time} \\
\end{split}
\end{equation}

If we now restrict to the typical scenario of just two neutrino states of definite masses,
${\ket{{\nu_1}}}$ and ${\ket{{\nu_2}}}$, then the relevant unitary mixing matrix is simply a
${2\times 2}$ submatrix of the general mixing matrix ${U}$ \cite{Kayser-2001}. As a result,
the transition probability for the neutrinos to ``oscillate'' from a given flavor state, say
${\ket{\nu_{\mu}(0)}}$, to another flavor state, say ${\ket{\nu_e(t)}}$, is given by
\begin{equation}
\begin{split}
P_{\nu_{\mu}\rightarrow\nu_e}(E,\,L)&=|\langle\nu_e(0)\ket{\nu_{\mu}(t)}|^2 \\
&=4|U_{\mu 2}|^2|U_{e 2}|^2\sin^2\left(\frac{\Delta m^2}{\,4E\,}L\right)\!,\label{transition} \\
\end{split}
\end{equation}
where ${\Delta m^2}$ ${\equiv}$ ${m_2^2-m_1^2}$ ${>0}$
is the difference in the squares of the two masses.
Now, in terms of the mixing angle ${\theta}$, the quantity ${4|U_{\mu 2}|^2|U_{e 2}|^2}$ simply
turns out to be equal to ${\sin^22\theta}$ (cf. \cite{Kayser-2001}). Using this, and a
trigonometric
identity, the transition probability can finally be recast in the following perspicuous form:
\begin{equation}
P_{\nu_{\mu}\rightarrow\nu_e}(E,\,L)=\frac{1}{2}\sin^22\theta
\left[1-\cos\left(\frac{\Delta m^2}{\,2E\,}L\right)
\right]\!,\label{final-prob}
\end{equation}
where, incidentally (due to (\ref{momen})),
\begin{equation}
\frac{\Delta m^2}{\,2E\,}\equiv \Delta p \equiv p_1-p_2\,.\label{recog}
\end{equation}
From this transition probability it is clear that the experimental observability of neutrino
flavor oscillations is essentially determined by the quantum phase
\begin{equation}
\Phi:=2\pi\,\frac{L\,}{L_O}\,,\label{quan-phase}
\end{equation}
where
\begin{equation}
L_O(E,\,m):=\frac{2\,\pi}{\Delta p}=\frac{4\pi E}{\,\Delta m^2}\label{osci-length}
\end{equation}
is the energy-dependent oscillation length. In particular, flavor changes would be observable
whenever the propagation distance of the neutrinos, ${L}$, is of the order of the oscillation
length, ${L_O}$. Therefore, in what follows, it would suffice to concentrate on these two
variables.

\section{Planck scale corrections to the oscillation length}

Now, let us assume that the standard theory reviewed above remains essentially valid for
ultra-high-energy neutrinos. It is then natural to wonder how the theory is generalized by
approaches to quantum gravity that give rise to modifications of the form (\ref{first}) of the
standard dispersion relation ${E^2=p^2+m^2}$. The question can be easily answered
by replacing the approximation (\ref{momen}) by
\begin{equation}
p_j\approx E-\frac{\,m^2_j\,}{2E\,}-\frac{\,\xi^{(1)}_j}{2}\frac{E^2}{\cs mP}
-\left(\frac{\,\left(\xi^{(1)}_j\right)^2}{8}+\frac{\,\xi^{(2)}_j}{2}\right)
\frac{E^3}{\css m2P}\,,\label{ptilde}
\end{equation}
which follows from the modified dispersion relations (\ref{first}) after the terms higher than
second-order in ${m_j}$ and ${\css m{-1}P}$, as well as the terms involving ${m_j^2/{\cs mP}}$,
are ignored. Here, following Refs.\cite{Alfaro-2000,Alfaro-2002,Lambiase-2003}, we have assumed
that the coefficients
${\xi^{(1)}}$ and ${\xi^{(2)}}$ in the expansion (\ref{first}) may in general depend on the
neutrino flavor (and hence on flavor-mixing, in accordance with Eq.(\ref{flavour})). The
corresponding modified oscillation length, analogous to (\ref{osci-length}), is then given by
\begin{equation}
{\widetilde{L}}_O(E,\,m,\,\xi^{(1)},\,\xi^{(2)}):=
\frac{2\,\pi}{\Delta p}=\frac{2\,\pi}{\frac{\,\Delta m^2}{\,2E\,}+\Delta q\,}\,,
\label{mod-osci}
\end{equation}
where we have set
\begin{equation}
\Delta q \equiv \frac{\,\Delta \xi^{(1)}}{2}\frac{E^2}{\cs mP}+\left(\frac{\,\Delta\!\!
\left(\xi^{(1)}_{}\right)^2}{8}+
\frac{\Delta \xi^{(2)}}{2}\right)\frac{E^3}{\css m2P}\,,\label{delta-q}
\end{equation}
along with 
\begin{equation}
\begin{split}
&\Delta \xi^{(1)} \equiv \xi^{(1)}_2-\xi^{(1)}_1, \\
&\Delta\!\!\left(\xi^{(1)}_{}\right)^2 \equiv
\left(\xi^{(1)}_2\right)^2-\left(\xi^{(1)}_1\right)^2\!,\;{\rm and} \\
&\Delta \xi^{(2)} \equiv \xi^{(2)}_2-\xi^{(2)}_1. \\
\end{split}
\end{equation}
This is clearly an odd result. It implies that, according to various approaches to quantum
gravity that lead to the modifications such as (\ref{first}), flavor oscillations can occur
even for neutrinos with negligible masses, or for massive neutrinos with degenerate
mass-eigenstates \cite{Lambiase-2003}. The corresponding purely quantum gravity induced
oscillation length---in either of the two cases---would be
\begin{equation}
L_{QG}(E,\,\xi^{(1)},\,\xi^{(2)})=\frac{2\,\pi}{\Delta q}\,,\label{qg-osci}
\end{equation}
where ${\Delta q}$ is again given by (\ref{delta-q}).
Of course, this curious conclusion depends on the above assumption that the coefficients
${\xi^{(n)}}$ may in general be flavor-dependent.

It is possible to avoid this conclusion altogether, however, within a different approach. Note
that the modification (\ref{mod-osci}) of the oscillation length arises from the replacement
of the Lorentz-invariant quantum phase factor (\ref{phase}) by a generalized factor with ${p_j}$
given by (\ref{ptilde}). There are at least two possible interpretations of this generalized
factor, depending on how
the Lorentz invariance is treated in a given approach to quantum gravity: Either the relativity
of inertial frames is taken as effectively broken in a theory, allowing the existence of a
privileged frame, or it is taken as preserved, but Lorentz transformations are made to act
non-linearly on the energy and momentum eigenstates of the theory. In the former case the
conservations of energy and momentum are assumed to remain linear, whereas in the latter case
they are deemed to be non-linear, affecting any experimental analysis \cite{Smolin-2003}.
There is, however, a third possibility in which the relativity of inertial frames is indeed
generalized at the fundamental level, but {\it without} permitting a privileged reference
frame, and {\it without} compromising the linearity of the conservations of energy and
momentum. A theory incorporating this largely unappreciated possibility has been proposed in
Ref.\cite{Christian-2004}, which gives rise to the following generalization of the special
relativistic dispersion relation:
\begin{equation}
p^2+m^2=E^2\left[1-\frac{\left(E-m\right)^2}{\css m2P}\right]\!.\label{exact}
\end{equation}

There are several features of this generalization that are worth bringing out. To begin with,
unlike the approximate nature of (\ref{first}), the above expression is {\it exact}, with no
other but a quadratic suppression by the Planck energy. Moreover, as it is supposed to be a
part of truly fundamental theory replacing special relativity at the Planck scale, there are
no parameters in the expression to be adjusted. And yet, independently of the Planck scale,
in the rest frame of the particle it duly reproduces ${E=mc^2}$. What is more, the expression
leads to a Planck scale induced modification of the neutrino flavor oscillations that
turns out to be quite interesting in its own right. To appreciate this,
note that according to (\ref{exact}), for ${m\ll E}$, and up to only the quadratic suppression
by the Planck energy, the modified momentum of a neutrino of mass ${m_j}$ can be approximated as
\begin{equation}
p_j\approx E-\frac{m_j^2}{2E}+\frac{E^2}{\css m2P}\,m_j\,,\label{mod-mom}
\end{equation}
provided the terms involving the ratio ${m_j^2/{\cs mP}}$ are also deemed negligible.
The corresponding modified oscillation length, analogous to (\ref{mod-osci}), is then given by
\begin{equation}
L^{'}_O(E,\,m):=\frac{2\,\pi}{\Delta p}=\frac{2\,\pi}{\frac{1}{2E}\,\Delta m^2\,-\,
\frac{E^2}{\,\css m2P}\,\Delta m}\,,\label{r-osci-r}
\end{equation}
where ${\Delta m^2 \equiv m^2_2-m^2_1}$ as before, and ${\Delta m \equiv m_2-m_1}$. For
neutrinos with energy much smaller than the Planck energy this generalized oscillation length
clearly reduces to the standard expression (\ref{osci-length}), whereas for
neutrinos with energy approaching the Planck energy the second term in the denominator of
(\ref{r-osci-r}) dominates, yielding
\begin{equation}
L^{'}_O(E,\,m)\rightarrow L_{RR}(E,\,m):=\frac{2\,\pi}{\Delta r}:=
\frac{-2\,\pi\,}{\Delta m}\frac{\,\css m2P}{E^2}\,.\label{pure-rr-osci}
\end{equation}
More significantly, and in sharp contrast with the peculiar implications of (\ref{mod-osci})
discussed above, the generalized length (\ref{r-osci-r}) implies (quite sensibly) that there
would be no flavor oscillations for massless particles, or for massive neutrinos with
degenerate mass-eigenstates.

\section{Observability of the Planck scale induced oscillations}

Let us now address the question of experimental distinguishability of the two modified
oscillation lengths above, (\ref{r-osci-r}) and (\ref{mod-osci}), from each other, and from
their special relativistic counterpart, (\ref{osci-length}). Beginning with the length
(\ref{r-osci-r}), it is clear that the Planck scale induced modification would become
significant in this case when
\begin{equation}
\frac{\,\Delta m^2\,}{2E}\sim\frac{E^2}{\,\css m2P}\,\Delta m\,,\label{sim}
\end{equation}
or, equivalently, when
\begin{equation}
L_O\equiv\frac{4\pi E}{\,\Delta m^2}\sim L_{RR}\equiv
\left\vert\frac{2\,\pi}{\,\Delta r\,}\right\vert=
\frac{\,2\,\pi\,}{\Delta m}\frac{\,\css m2P}{E^2}\,.\label{RR-equiv}
\end{equation}
Of course, for those neutrinos of energy such that (\ref{sim}) is an {\it exact} equality,
flavor oscillations would be washed out, providing a distinctive signature for the modified
relation (\ref{exact}). More generally, the energy necessary to reveal deviations from the
standard flavor oscillations would depend on the sensitivity with which the value of the mass
splittings can be inferred in a given experiment.
In experiments performed to date, this sensitivity
ranges from ${\Delta m^2\sim 1}$ ${({\rm eV})^2}$, for short-baseline accelerator neutrinos,
to ${\Delta m^2\sim 10^{-11}}$ ${({\rm eV})^2}$, for solar neutrinos \cite{Kayser-2001}. These
values can be easily calculated by noting from (\ref{final-prob}) that neutrino flavors
oscillate as a function of ${L/E}$, and that ${\Delta m^2}$ reach for such oscillations is
inversely proportional to this ratio:
\begin{equation}
\Delta m^2\sim\frac{4\,\pi\,E}{L}\,.\label{infact}
\end{equation}
Substituting this reach into the condition (\ref{sim}) (along with the assumption
${m_2\gg m_1}$) yields 
\begin{equation}
L\sim\frac{\,\pi\,\css m4P}{E^5}\,.\label{L-right}
\end{equation}
This, then, is the necessary constraint between the neutrino energy ${E}$ and its propagation
distance ${L}$, for detecting significant deviations from the standard
flavor oscillations. For example, it can be easily calculated from this condition that the
Planck scale deviations in the oscillation length, induced by the generalization (\ref{exact}),
would be either observable, or can be ruled out, for neutrinos of energy ${E\sim 10^{17}}$ eV,
provided that they have originated from sources located at some ${10^{5}}$ light-years away from
the detector. According to (\ref{infact}), the corresponding confidence in the mass splittings
${\Delta m^2}$ would then be of the order of ${10^{-10}}$ ${({\rm eV})^2}$, which is comparable
to that achieved for the solar neutrinos \cite{Kayser-2001}.

Turning now to the modified oscillation length (\ref{mod-osci}), it is clear, once again, that
the Planck scale induced modification would become significant in this case when
\begin{equation}
\frac{\,\Delta m^2\,}{2E}\sim\Delta q\,,\label{again-sim}
\end{equation}
with ${\Delta q}$ given by (\ref{delta-q}), or, equivalently, when
\begin{equation}
L_O\equiv\frac{4\pi E}{\,\Delta m^2}\sim L_{QG}\equiv
\frac{2\,\pi}{\Delta q\,}\,.\label{equiv-again}
\end{equation}
Now, as we noted after the definition (\ref{osci-length}), for flavor changes to be
observable at all, the propagation distance ${L}$ of the neutrinos must be about the same
size as their oscillation length ${L_O}$. Moreover, it is easy to infer from (\ref{ptilde})
and (\ref{delta-q}) that ${\Delta q}$ is an expansion of the form
\begin{equation}
\Delta q \approx \sum_{n=1} f_n \frac{E^{n+1}}{\css mnP}\,,\label{expansion}
\end{equation}
with ${f_n}$ being the functions of various splittings such as ${\Delta \xi^{(n)}}$, even when
the order of suppression is kept arbitrary. Consequently, putting all of the above observations
together, it is easy to see that the Planck scale deviations from the standard oscillation
length would become significant in this case, for {\it each} order of suppression by the Planck
energy, when
\begin{equation}
L\sim\frac{2\,\pi}{\,f_n\,}\,\frac{\css mnP}{E^{n+1}}.\label{signi-sim}
\end{equation}

Now, up to linear suppression by the Planck energy, ${f_1\equiv {\Delta \xi^{(1)}}/2}$, and
the existing data from long-baseline accelerator neutrinos already imply that the splittings
${\Delta \xi^{(1)}}$ cannot be greater than ${10^{-4}}$ (using values from
Ref.\cite{Kayser-2001}, for such terrestrial neutrinos ${E\sim 10}$ GeV and ${L\sim 10^{3}}$
km, which, upon substitutions in (\ref{signi-sim}), gives ${\Delta \xi^{(1)}\lesssim 10^{-4}}$).
Unfortunately, since quantum interferences
can only reveal {\it relative} phases of the interfering amplitudes, flavor oscillations can
only be sensitive to the splittings such as ${\Delta \xi^{(1)}}$, and not to the underlying
individual coefficients such as ${\xi^{(1)}_1}$ or ${\xi^{(1)}_2}$. On the other hand, provided
cosmogenic neutrinos are at our disposal, remarkably strong bounds on the splittings
${\Delta \xi^{(n)}}$---for at least up to the {\it seventh} order of suppression---can be
obtained with reasonable neutrino energies.

To appreciate this, suppose we demand that the splittings ${\Delta \xi^{(n)}}$ be probed
${\sigma}$ times as deeply as the mass splittings ${\sqrt{\Delta m^2\,}}$; that is, suppose that
\begin{equation}
\Delta \xi^{(n)}\sim\sigma\,\sqrt{\frac{\Delta m^2}{m^2_e}}\sim\sigma\,\sqrt{
\frac{4\,\pi\,E}{m^2_e\,L}}\,,
\label{sigmatimes}
\end{equation}
where ${m_e}$ is the electron mass (as standardly used in the neutrino experiments), and the
last relation follows from (\ref{infact}). Then, assuming that {\it all} but the ${n^{th}}$
order coefficient ${\xi_j^{(n)}}$ in the expansion (\ref{first}) are negligible (i.e., reducing
${f_n}$ to ${\Delta\xi^{(n)}/2\,}$), the condition (\ref{signi-sim}) becomes
\begin{equation}
L\sim\frac{4\pi\,m^2_e}{\sigma^2}\,\frac{\;\css m{2n}P}{E^{2n+3}}\,,\label{L-E-condi}
\end{equation}
which specifies the constraint between the energy ${E}$ and propagation distance ${L}$ of
neutrinos for probing the ${n^{th}}$ order correction due to the modified dispersion relations
(\ref{first}). For example, for the case of quadratic suppression by the Planck energy
(${n=2}$), with ${\sigma}$ as stringent as ${10^{-10}}$, condition (\ref{L-E-condi}) dictates
the values ${E\sim 10^{16}}$ eV and ${L\sim 10^{10}}$ light-years, giving ${\Delta m^2}$ reach
${\sim 10^{-16}}$ ${(\rm eV)^2}$ and ${\Delta\xi^{(2)}}$ reach ${\sim 10^{-24}}$. These
remarkable values may seem incredible, but it can be easily checked that the higher-order terms
neglected in the expansion (\ref{ptilde}) (with ${n\equiv 2}$) continue to remain negligible
even with these minute values for ${\Delta m^2}$ and ${\Delta\xi^{(2)}}$. On the other hand,
neither condition (\ref{L-E-condi}) nor condition (\ref{L-right}) necessitates ${\Delta m^2}$
to be as small as ${10^{-16}}$ ${(\rm eV)^2}$ for the quantum gravity effects to be observable.
In fact, it can be as large as ${10^{-4}}$ ${(\rm eV)^2}$, or larger, for reasonable values of
${E}$ and ${L}$. For instance, for ${E\sim 10^{19}}$ eV and ${L\sim 10}$ light-years, yielding
${\Delta m^2\sim 10^{-4}}$ ${(\rm eV)^2}$, the condition (\ref{L-E-condi}) (or
(\ref{signi-sim})), with ${n=2}$, leads to ${\Delta\xi^{(2)}}$ reach as deep as
${\sim 10^{-24}}$.

Alternatively, one may wish to relinquish this high sensitivity of ${\Delta\xi^{(n)}}$ for the
sake of probing deviations from ${E^2=p^2+m^2}$ that are suppressed by ${\cs mP}$ at a much
higher order. As an extreme example, one can imagine a model in which quantum gravity induced
deviations become significant only at the {\it seventh} order of suppression by the Planck
energy. Then, with ${\sigma}$-setting as high as ${10^{+8}}$, condition (\ref{L-E-condi}), with
${n=7}$, dictates the values of ${E\sim 10^{21}}$ eV and ${L\sim 10^{9}}$ light-years, giving
${\Delta m^2}$ reach ${\sim 10^{-10}}$ ${(\rm eV)^2}$ and ${\Delta\xi^{(7)}}$ reach
${\sim 10^{-3}}$. Thus, flavor oscillating neutrinos with energy in the range of ZeV---provided
they have originated at a cosmological distance of some ${10^{9}}$ light-years---are
capable of probing the effects of quantum gravity that may be as minute as, say,
{\it septically} suppressed by the Planck energy.

\section{Prospects for observations in the median future}

It is clear from the above discussion that, at least in principle, ultra-high-energy neutrinos
of cosmic origin can serve as a remarkably sensitive probe of the Planck regime. In fact,
within the next decade, thanks to the surge of progressively larger neutrino detectors, even
in practice this intriguing possibility may become reality.

To be sure, as yet no extraterrestrial neutrino of energy greater than a few tens of MeV has
been observed by any of the existing detectors. On the other hand, as we saw above, the energy
range of neutrinos necessary to probe the Planck regime may be between
10 PeV to ZeV, with the corresponding propagation distance ranging from ${10^{5}}$ light-years
to ${10^{10}}$ light-years. This gap in energy is quite likely to be closed within a decade,
however, for some of the neutrino detectors presently under construction are indeed expected
to detect significant fluxes of neutrinos of energy up to and beyond ZeV, originating from
giant astrophysical sources such as active galactic nuclei (AGN) and gamma ray bursters (GRBs).
Moreover, the atmospheric neutrino background of Earth happens to have such a large spectral
index that, for neutrino detections in the energy region above 10 PeV, it is unlikely to
significantly obscure the cosmogenic neutrino flux, regardless of its source \cite{Halzen-2001,
Sasaki-2002}. Furthermore, since neutrinos are neutral and stable particles, they point back to
their sources, thereby providing vital information about their propagation distances (in terms
of cosmological redshifts; cf., \cite{Lunardini-2001}). Consequently, the next
generation of neutrino detectors may well be able to witness quantum gravity effects, at
least of the kind discussed above.

There are a variety of different cosmogenic neutrino detectors under construction at
present, or planned to be operational within the next decade (for an introductory survey, see,
e.g., Ref.\cite{Spiering-2003}). For instance, IceCube---a cubic kilometer size detector
under construction at the South Pole \cite{IceCube}---is expected to be fully operational by
the year 2009. It is optimized for the energy range of ${10^{11}}$ to ${10^{18}}$ eV, but will
be sensitive to energies up to ${10^{20}}$ eV \cite{Halzen-2001}. Two other
projects for large neutrino detectors are under construction in the middle of the
Mediterranean, namely, ANTARES \cite{ANTARES} and NESTOR \cite{NESTOR}, not to mention the
existing underwater detector located at a depth of 1100 meters in the Siberian Lake Baikal
\cite{Baikal}. All three of these projects envisage larger cubic kilometer size extensions,
with operational capacities comparable to those of IceCube, or its predecessor AMANDA
\cite{AMANDA}.

The typical optimized energy range of the above underwater or under-ice detectors is from TeV
to PeV, with sensitivity extendible up to ${\sim 10}$ EeV. Recently, there has been a renewed
impetus for constructing detectors that can study neutrinos of super-EeV energies, with
sensitivity of some of them ranging well beyond the ZeV scale. These include Auger \cite{AUGER},
SAUND \cite{SAUND}, RICE \cite{RICE}, ANITA \cite{ANITA}, GLUE \cite{GLUE}, OWL \cite{OWL},
and ASHRA \cite{ASHRA,Sasaki-2002}.
For example, ANITA experiment is planned for a 30 day balloon flight over
Antarctica as early as in 2007, and expected to collect a significant flux of ultra-high-energy
neutrinos. Of much wider scope is a relatively new proposal, ASHRA, which purports to
conduct an ``entirely all-sky survey'' by a telescopic array---based near three mountain
peaks in the Big Island of Hawaii---that will explore neutrinos of energy up to,
and beyond, several ZeV.

The above list of neutrino detectors is by no means exhaustive, nor does it purport to do
justice to their true potential. But it does give a good indication of the intense efforts
that are underway for detecting the ultra-high-energy cosmogenic neutrinos. In general,
however, these efforts do have to face a major difficulty. Due to their astronomically long
propagation lengths, formidable challenges lay ahead in collecting statistically significant
flux of such cosmogenic neutrinos, especially of those with energies above the threshold of
PeV. In spite of these potential difficulties, however, many experimental collaborations are
expecting to detect significant fluxes of cosmogenic neutrinos within the next decade. For
example, the ASHRA Collaboration mentioned above is expecting to detect more than ${1300}$
events per year---at the threshold energy of ${10^{19}}$ eV---once their telescopic array
becomes fully operational. Apart from the increasingly innovative detection techniques, such
optimistic expectations stem largely from the theoretical
estimates obtained so far for the initial flux
of neutrinos from the cosmologically distant sources \cite{Engel-2001, Kalashev-2002}.

Ultra-high-energy neutrinos from sources such as AGN and GRBs are usually thought to be produced
as secondaries of cosmic ray protons interacting with ambient matter and photon fields (for a
review, see, e.g., Ref.\cite{Athar-2002}). Such proton-proton and proton-photon interactions
produce neutral and charged pions, which, in turn, decay into neutrinos via the chain:
${\pi^+\rightarrow\mu^+\nu_{\mu}\rightarrow e^+\,\nu_e\,\overline{\nu}_{\mu}\,\nu_{\mu}\,}$.
From inception, these interactions have been thought to provide a ``guaranteed'' source of
cosmogenic
neutrinos \cite{Berezinsky-1969}. Moreover, although the absolute flux of the different flavor
states of such neutrinos is presently unknown, the above decay chain strongly suggests their
relative flux ratios ${\phi^S_{\nu_e}\!:\phi^S_{\nu_{\mu}}\!:\phi^S_{\nu_{\tau}}}$ at the source
to be ${\frac{1}{3}:\frac{2}{3}:\frac{0}{3}}$. (Here the flux of tau neutrinos---produced in the
decay chain of charmed mesons in the same reaction---is thought to be suppressed well below the
order of ${10^{-3}}$). Implications of these initial flux ratios (in the literature usually
referred to as ``the standard ratios'') have been studied extensively in recent years,
especially in the context of neutrino oscillations
\cite{Athar-2000, Ahluwalia-2001,Beacom-2003}. However, these studies are based on a simplifying
assumption that the propagation distance ${L}$ of such neutrinos is astronomically large
compared to the oscillation length ${L_O}$; or, equivalently, that the relative quantum phase
(\ref{quan-phase}) acquired by the propagating mass-eigenstates is oscillating very rapidly.
Consequently, in the scenarios considered in these studies, the neutrinos arriving at a
terrestrial detector would be in an incoherent mixture of mass-eigenstates, with the above
initial flux ratios reduced to ${\phi^D_{\nu_e}\!:\phi^D_{\nu_{\mu}}\!:\phi^D_{\nu_{\tau}}\sim
\frac{1}{3}:\frac{1}{3}:\frac{1}{3}}$.

By contrast, the relation ${L\sim L_O}$ is intrinsic to both observability conditions,
(\ref{L-right}) and (\ref{L-E-condi}), derived above, and hence the quantum coherence will be
maintained for neutrinos in these cases, throughout their long journey through the intervening
vacuum. This coherence would then be reflected in the different flavor fluxes
registered at the terrestrial detectors, which can be estimated as
\begin{equation}
\phi^D_{\nu_{\beta}}\;=\!\!\sum_{\alpha\,=\,e,\,\mu,\,\tau}\!\!P_{\alpha\beta}(E)\;
\phi^S_{\nu_{\alpha}}\,,\label{at-detector}
\end{equation}
where
\begin{equation}
\begin{split}
P_{\alpha\beta}(E):=&\;P_{\nu_{\alpha}\rightarrow\,\nu_{\beta}}(E) \\
=&\;\delta_{\alpha\beta}-\!\!\sum_{j\not=k}U^*_{\alpha j}U_{\beta j}
U_{\alpha k}U^*_{\beta k}\!\left(\!1-e^{-i\frac{\Delta m^2_{jk}}{2E}L}\!\right) \\
\end{split}
\label{detect-prob}
\end{equation}
are the three-flavor transition probabilities, analogous to the two-flavor ones given by
(\ref{transition}). In practice, however, these transition probabilities would have to be
averaged over the redshift distribution of neutrino sources \cite{Lunardini-2001}, in order to
incorporate the effects of their evolution with respect to the corresponding cosmological epoch
${z}$:
\begin{equation}
\overline{P_{\alpha\beta}}(E)=\frac{\int_0^{z_{max}}P_{\alpha\beta}(E,\,z)\;
f_{\nu_{\alpha}}(z)\,dz}{\int_0^{z_{max}}f_{\nu_{\alpha}}(z)\,dz}\,,\label{distrib}
\end{equation}
where ${P_{\alpha\beta}(E,\,z)}$---which can be obtained by replacing ${E}$ with ${(1+z)E}$ in
the expression given by (\ref{detect-prob})---are the transition probabilities for neutrinos
produced in the epoch ${z}$, and
${f_{\nu_{\alpha}}(z):=\frac{\;d\phi^S_{\nu_{\alpha}}\!(z)}{dz}}$ represents
the redshift distribution of cosmogenic sources producing the neutrinos ${\nu_{\alpha}}$
(examples of ${f_{\nu_{\alpha}}(z)}$ have been studied in Ref.\cite{Lunardini-2001}).

It is clear from the above discussion that, given the standard flux ratios
${\phi^S_{\nu_e}\!:\phi^S_{\nu_{\mu}}\!:\phi^S_{\nu_{\tau}}\sim
\frac{1}{3}:\frac{2}{3}:\frac{0}{3}}$ for active neutrinos being produced at a cosmogenic
source, the flux ratios ${\phi^D_{\nu_e}\!:\phi^D_{\nu_{\mu}}\!:\phi^D_{\nu_{\tau}}}$ to be
registered at a terrestrial detector can be estimated using the relation (\ref{at-detector}),
which in turn depends on the oscillation length (\ref{osci-length}) via the transition
probabilities (\ref{detect-prob}), or (\ref{distrib}). These flux ratios would be different,
however, if, instead of (\ref{osci-length}), either the oscillation length (\ref{mod-osci}), or
(\ref{r-osci-r}), is respected by nature, provided the corresponding observability condition
(\ref{L-E-condi}), or (\ref{L-right}), is satisfied. In other words, the observed ratios would
be different for predictions based on the dispersion relations (\ref{first}), or (\ref{exact}),
from those based on ${E^2=p^2+m^2}$.

\section{Concluding remarks}

We have explored implications of some theories of quantum gravity for neutrino flavor
oscillations, within the context of modified dispersion relations (\ref{first}) and
(\ref{exact}). In particular, we have obtained expressions for Planck-scale-induced
deviations from the standard oscillation length, implied by these modified dispersion
relations, as functions of neutrino mass, energy, and propagation length. These expressions
suggest that, in general, even those deviations that are quadratically suppressed by the Planck
energy may become observable for ultra-high-energy neutrinos, originating at cosmological
distances. In fact, for neutrinos in the highest energy range of EeV to ZeV, deviations that are
as minute as up to {\it septically} suppressed by the Planck energy may become observable.
As a result---and since the next generation of neutrino detectors are poised to receive
sizable flux of ultra-high-energy cosmogenic neutrinos, the chances of empirically
verifying (or ruling out)
such minute quantum gravity effects---within a decade---are quite promising.

\begin{acknowledgments}

I would like to thank Ted Jacobson for detecting an error in the references, and
D. V. Ahluwalia-Khalilova and Subir Sarkar for their comments on the manuscript.

\end{acknowledgments}


\begin{thebibliography}{}

\bibitem{Rovelli-2000}C. Rovelli, ``Notes for a
brief history of quantum gravity'' [arXiv:gr-qc/0006061].

\bibitem{Amelino-Camelia-2005}G. Amelino-Camelia and J. Kowalski-Glikman (eds.), ``Quantum
Gravity Phenomenology'', Lecture Notes in Physics (Springer-Verlag, Heidelberg, 2005).

\bibitem{Samuel-1989}V.A. Kosteleck\'y and S. Samuel,  Phys. Rev. D {\bf 39}, 683 (1989).

\bibitem{Ellis-2000a}J. Ellis, N.E. Mavromatos, and D.V. Nanopoulos, Phys. Rev. D {\bf 61},
027503 (2000).

\bibitem{Pullin-1999}R. Gambini and J. Pullin, Phys. Rev. D {\bf 59}, 124021 (1999).

\bibitem{Alfaro-2000}J. Alfaro, H.A. Morales-T\'ecotl, and L.F. Urrutia, Phys. Rev. Lett.
{\bf 84}, 2318 (2000).

\bibitem{Alfaro-2002}J. Alfaro, H.A. Morales-T\'ecotl, and L.F. Urrutia, Phys. Rev. D
{\bf 66}, 124006 (2002).

\bibitem{Lehnert-2003}R. Lehnert, Phys. Rev. D {\bf 68}, 085003 (2003).

\bibitem{Bertolami-2004}O. Bertolami, R. Lehnert, R. Potting, and A. Ribeiro, Phys. Rev. D
{\bf 69}, 083513 (2004).

\bibitem{Smolin-2003}L. Smolin, ``How far are we from the quantum theory of gravity''
[arXiv:hep-th/0303185].

\bibitem{Amelino-Camelia-1998}G. Amelino-Camelia, J. Ellis, N.E. Mavromatos,
D.V. Nanopoulos, and S. Sarkar, Nature {\bf 393}, 763 (1998).

\bibitem{Jacobson-2003}T. Jacobson, S. Liberati, and D. Mattingly, Nature {\bf 424},
1019 (2003).

\bibitem{Jacobson-2003b}T. Jacobson, S. Liberati, and D. Mattingly, Phys. Rev. D {\bf 67},
124011 (2003).

\bibitem{Stecker-2004}T. Jacobson, S. Liberati, D. Mattingly, and F.W. Stecker, Phys.
Rev. Lett. {\bf 93}, 021101 (2004).

\bibitem{Jackiw-1990}S.M. Carroll, G.B. Field, and R. Jackiw, Phys. Rev. D {\bf 41}, 1231
(1990).

\bibitem{Mewes-2001}V.A. Kosteleck\'y and M. Mewes, Phys. Rev. Lett. {\bf 87}, 251304 (2001).

\bibitem{Amelino-Camelia-2003}G. Amelino-Camelia, Int.J.Mod.Phys. D {\bf 12}, 1633 (2003).

\bibitem{ANTARES}T. Montaruli [ANTARES Collaboration], Nucl. Phys. Proc. Suppl.
{\bf 110}, 513 (2002).

\bibitem{Jacobson-2004}T. Jacobson, S. Liberati, and D. Mattingly, ``Astrophysical bounds on
Planck suppressed Lorentz violation'' [arXiv:hep-ph/0407370].

\bibitem{Lambiase-2003}G. Lambiase, Mod. Phys. Lett. A {\bf 18}, 23 (2003); Class. Quant. Grav.
{\bf 20}, 4213 (2003).

\bibitem{Brustein-2002}R. Brustein, D. Eichler, and S. Foffa, Phys. Rev. D {\bf 65},
105006 (2002).

\bibitem{Ellis-2000}J. Ellis, N.E. Mavromatos, D.V. Nanopoulos, and G. Volkov,
Gen. Rel. Grav. {\bf 32}, 1777 (2000).

\bibitem{Bertolam-2000}O. Bertolami and C.S. Carvalho, Phys. Rev. D {\bf 61}, 103002 (2000).

\bibitem{Barger-2000}V. Barger, S. Pakvasa, T.J. Weiler, and K. Whisnant, Phys. Rev. Lett.
{\bf 85}, 5055 (2000).

\bibitem{Bahcall-2002}J.N. Bahcall, V. Barger, and D. Marfatia, Phys. Lett. B {\bf 534}, 120
(2002).

\bibitem{Choubey-2003}S. Choubey and S.F. King, Phys. Rev. D {\bf 67}, 073005 (2003).

\bibitem{Mewes-2004}V.A. Kosteleck\'y and M. Mewes, Phys. Rev. D {\bf 70}, 031902 (2004).

\bibitem{IceCube}J. Ahrens {\it et al.} [IceCube Collaboration], Astropart. Phys. {\bf 20},
507 (2004).

\bibitem{ANITA}ANITA, http://www.ps.uci.edu/${\!\sim}$barwick/anitaprop.pdf

\bibitem{Christian-2004}J. Christian, Int. J. Mod. Phys. D {\bf 13}, 1037 (2004)
[arXiv:gr-qc/0308028].

\bibitem{Kayser-2001}B. Kayser, ``Neutrino mass, mixing, and oscillation''
[arXiv:hep-ph/0104147].

\bibitem{Beuthe-2003}M. Beuthe, Physics Reports {\bf 375}, 105 (2003).

\bibitem{Stodolsky-1998}L. Stodolsky, Phys. Rev. D {\bf 58}, 036006 (1998).

\bibitem{Giunti-2004}C. Giunti, Am. J. Phys. {\bf 72}, 699 (2004).

\bibitem{Halzen-2001}J. Alvarez-Muniz and F. Halzen, Phys. Rev. D {\bf 63}, 037302 (2001).

\bibitem{Sasaki-2002}M. Sasaki, Y. Asaoka, and M. Jobashi, Astropart. Phys. {\bf 19}, 37 (2003).

\bibitem{Lunardini-2001}C. Lunardini and A. Yu. Smirnov, Phys. Rev. D {\bf 64}, 073006 (2001).

\bibitem{Spiering-2003}C. Spiering, J. Phys. G {\bf 29}, 843 (2003); Nucl. Phys. Proc. Suppl.
{\bf 125}, 1 (2003).

\bibitem{NESTOR}P. Grieder {\it et al.} [NESTOR Collaboration], Nucl. Phys. B (Proc. Suppl.)
{\bf 97}, 105 (2001).

\bibitem{Baikal}D. Domogatski {\it et al.} [Baikal Collaboration], Nucl. Phys. B (Proc. Suppl.)
{\bf 118}, 363 (2003).

\bibitem{AMANDA}J. Ahrens {\it et al.} [AMANDA Collaboration], Phys. Rev. Lett. {\bf 92},
071102 (2004).

\bibitem{AUGER}Pierre Auger Observatory, http://www.auger.org/ 

\bibitem{SAUND}J. Vandenbroucke {\it et al.} [SAUND Collaboration], ``Experimental study of
acoustic ultra-high-energy neutrino detection'' [arXiv:astro-ph/0406105].

\bibitem{RICE}I. Kravchenko {\it et al.} [RICE Collaboration], Astropart. Phys. {\bf 20}, 195
(2003).

\bibitem{GLUE}P.W. Groham {\it et al.} [GLUE Collaboration], Phys. Rev. Lett. {\bf 93}, 041101
(2004).

\bibitem{OWL}F.W. Stecker {\it et al.} [OWL Collaboration], Nucl. Phys. B (Proc. Suppl.)
{\bf 136}, 433 (2004).

\bibitem{ASHRA}ASHRA, http://www.icrr.u-tokyo.ac.jp/${\!\sim}$ashra/

\bibitem{Engel-2001}R. Engel, D. Seckel, and T. Stanev, Phys. Rev. D {\bf 64}, 093010 (2001).

\bibitem{Kalashev-2002}O.E. Kalashev, V.A. Kuzmin, D.V. Semikoz, and G. Sigl, Phys. Rev. D
{\bf 66}, 063004 (2002).

\bibitem{Athar-2002}H. Athar, ``Some aspects of neutrino astrophysics'', arXiv:hep-ph/0212387.

\bibitem{Berezinsky-1969}V.S. Berezinsky and G.T. Zatsepin, Phys. Lett. B {\bf 28}, 423 (1969).

\bibitem{Athar-2000}H. Athar, M. Je\.zabek, and O. Yasuda, Phys. Rev. D {\bf 62}, 103007
(2000). 

\bibitem{Ahluwalia-2001}D.V. Ahluwalia, Mod. Phys. Lett. A {\bf 16}, 917 (2001).

\bibitem{Beacom-2003}J.F. Beacom, N.F. Bell, D. Hooper, S. Pakvasa, and T.J. Weiler,
 Phys. Rev. D {\bf 68}, 093005 (2003).

\end{thebibliography}
\end{document}